\documentclass[a4paper,12pt]{article}
\begin{document}
\begin{titlepage}
\begin{flushright}
\begin{tabular}{l}
UTHEP-508\\
RIKEN-TH-51\\
KEK-TH-1031\\
hep-th/0507263
\end{tabular}
\end{flushright}

\vspace{5mm}

\begin{center}
{\Large \bf Universality of Nonperturbative Effects\\ 
in $c<1$ Noncritical String Theory}
\baselineskip=24pt

\vspace{15mm}
\large
Nobuyuki Ishibashi,\footnote{ishibash@het.ph.tsukuba.ac.jp}\\
{\it Institute of Physics, University of Tsukuba,\\
     Tsukuba, Ibaraki 305-8571, Japan},\\
Tsunehide Kuroki,\footnote{kuroki@riken.jp}\\
{\it Theoretical Physics Laboratory, RIKEN,\\ 
     Wako 2-1, Saitama 351-0198, Japan}\\
and\\
Atsushi Yamaguchi,\footnote{ayamagu@post.kek.jp}\\
{\it High Energy Accelerator Research Organization (KEK),\\
Tsukuba, Ibaraki 305-0801, Japan }
\vspace{20mm}
\end{center}
\begin{abstract}
Nonperturbative effects in $c<1$ noncritical string theory 
are studied using the two-matrix model. 
Such effects are known to have the form fixed by the string equations but 
the numerical coefficients have not been known so far. 
Using the method proposed recently, we show that it is possible to determine 
the coefficients for $(p,q)$ string theory. 
We find that they are indeed finite in the double scaling limit 
and universal in the sense that they do not depend on the detailed structure 
of the potential of the two-matrix model. 
\end{abstract}
\end{titlepage}
\section{Introduction}
The nonperturbative effects in string theory reveal one of the most stringy 
features of string theory.  From the large-order behavior in perturbation series 
\cite{Shenker:1990uf}, or the calculation of the D-instanton effects \cite{Polchinski:1994fq}, 
one can see that the nonperturbative effects are of the form $\exp(-S_0/g_{\rm s})$, 
where $g_{\rm s}$ is the string coupling constant. 
These are quite different from the nonperturbative effects for a point particle theory which 
are of the form $\exp(-S_0/g^2)$ in terms of the coupling constant $g$. 
Studying such effects is important because we may be able to get some clues about the 
nonperturbative formulation of string theory by doing so. 
Noncritical string theory is a useful toy model for such an investigation. 
It possesses much fewer degrees of freedom compared with the critical one, and 
it can be nonperturbatively defined via a matrix model. 
Nevertheless it has many features in common with critical string theories and 
we can get insight about critical ones by studying the noncritical ones. 

Nonperturbative effects in noncritical string theory 
have been studied by many authors. 
Especially in \cite{David:1990sk,Eynard:1992sg,Fukuma:1996hj}, 
the value of $S_0$ is derived from the string equation, or as the action 
of solitonic excitations. 
More recently, nonperturbative effects 
in the $c=0$ noncritical string theory are analyzed concretely 
using the one-matrix model \cite{Hanada:2004im}. 
In \cite{Hanada:2004im}, the next to leading order contributions to the nonperturbative effects, which 
can be identified with the chemical potential of D-instantons,  
were computed by the method of orthogonal polynomials. 
It was shown that the nonperturbative effects up to this order 
are universal in the sense that these were independent of 
details of the matrix model potential. 
Since one cannot fix the exact value of the next to leading contribution from the 
string equation, it implies that the string equation fails to include some information of 
the matrix model. 
This result was further discussed and 
generalized in \cite{Sato:2004tz,Kawai:2004pj,deMelloKoch:2004en}.  

{}A natural question is 
whether the results in \cite{Hanada:2004im} can be generalized to other 
noncritical string theories. 
Since $(p,q)$ noncritical string theory is defined 
in a nonperturbative manner by taking an appropriate double scaling limit 
\cite{Brezin:1990rb,Gross:1989vs,Douglas:1989ve} 
of the two-matrix model \cite{Douglas:1989dd,Tada:1990kk,Tada:1991pa}, 
it is necessary to examine nonperturbative effects by using the two-matrix model. 
This is the problem we address in this paper. 

For this purpose, we first define the effective potential of the matrix 
eigenvalues as in \cite{Hanada:2004im}. 
The nonperturbative effects are due to the stationary points of this 
effective potential. 
In \cite{Kazakov:2004du} 
the leading order contributions to the nonperturbative effects 
are obtained from this effective potential for the two-matrix model 
and are shown to coincide 
with the results in the continuum approach given in \cite{Alexandrov:2003nn}.  
(Each stationary point corresponds to various ZZ-brane 
\cite{Zamolodchikov:2001ah}.) 
What we would like to do is to calculate the next to leading order contribution. 
In order to do this, we generalize the method proposed recently in \cite{Ishibashi:2005dh} 
to the two-matrix model case. 
We further prove that the result is universal in the double scaling limit 
as in the $c=0$ case.

This paper is organized as follows. 
In section 2, we define the effective potential of the matrix eigenvalues for 
the two-matrix model. The nonperturbative corrections to the free energy 
can be expressed in terms of this potential. 
In section 3, we summarize some of the known facts about the two-matrix model 
which will be used in section 4 and appendix B.  
In section 4, we calculate the nonperturbative effects up to the next leading order,  
and find that they are universal in the double scaling limit. 
Section 5 is devoted to conclusions and discussions.
In appendix A we present a result 
of the chemical potentials in the case of higher critical points 
of the one-matrix model in order to compare with our results for the two-matrix model.
In appendix B details of the computation of the denominator in the definition of 
the chemical potential are given.

\section{Nonperturbative effects in $c<1$ noncritical string theory}

The noncritical string theories for $c< 1$ can be
studied by using the two-matrix model,
\begin{equation}
\int {\rm d}{\bf X} {\rm d}{\bf Y} 
 \exp\left\{
            -\frac{N}{g}
            {\rm Tr}
              \left[
                    U({\bf X})+\widetilde{U}({\bf Y})-{\bf X}{\bf Y}
              \right]
     \right\}\,. \label{matrixintegral}
\end{equation}
Here, ${\bf X}, {\bf Y}$ are $N\times N$ hermitian matrices, 
and $U({\bf X}), \widetilde{U}({\bf Y})$
are polynomials of ${\bf X}, {\bf Y}$, respectively. 
This matrix integral 
can be expressed as an integral over the eigenvalues of ${\bf X},{\bf Y}$:
 \begin{equation}
\int \prod_{i=1}^N {\rm d}x_i {\rm d}y_i 
      \triangle(x)\triangle(y)
       \exp\left\{
                  -\frac{N}{g}
                   \sum_{i=1}^N \left[
                                      U(x_i)+\widetilde{U}(y_i)-x_i y_i
                                \right]
           \right\}\,,
\label{eigenvaluesintegral}
\end{equation}
where $\triangle(z)=\prod_{i>j}(z_i-z_j)$ 
is the Vendermonde determinant.

In order to study the nonperturbative
effects, it is convenient to define the effective potential
$V_{{\rm eff}}(x,y)$ of the matrix eigenvalues, as in the one-matrix 
case \cite{Hanada:2004im}. 
Picking up the $N$-th
eigenvalue $(x_N,y_N)$, we represent it as $(x,y)$. 
Then,
integrating over the other eigenvalues, we obtain
\begin{eqnarray}
&& \int\prod_{i=1}^{N-1} {\rm d}x_i^\prime {\rm d}y_i^\prime 
   \triangle(x^\prime)\triangle(y^\prime)
    \exp\left\{
               -\frac{N}{g}
                \sum_{i=1}^{N-1} 
                  \left[
                        U(x_i^\prime)+{\widetilde{U}}(y_i^\prime)
                        -x_i^\prime y_i^\prime
                  \right]
        \right\}
\nonumber\\
&&\qquad\qquad\qquad 
  \times \prod_{i=1}^{N-1} (x-x_i^\prime)(y-y_i^\prime) 
    \exp
     \left\{
            -\frac{N}{g} 
              \left[
                    U(x)+{\widetilde{U}}(y)-x y
              \right]
     \right\}\,. \label{boltzmann}
\end{eqnarray}
This quantity is regarded as the Boltzmann weight for $(x,y)$. 
Rewriting eq.(\ref{boltzmann}) in terms of $(N-1)\times (N-1)$
hermitian matrices ${\bf X}^\prime,{\bf Y}^\prime$, we can define $V_{\rm eff}(x,y)$ in the
following manner:
\begin{eqnarray}
&&\exp\left[
           -V_{{\rm eff}}(x,y)
     \right]\nonumber\\
&&=\frac
   {\int {\rm d}{\bf X}^\prime{\rm d}{\bf Y}^\prime
     \exp\left\{
                -\frac{N-1}{g^\prime} 
                {\rm Tr}
                  \left[
                        U({\bf X}^\prime)+\widetilde{U}({\bf Y}^\prime)
                        -{\bf X}^\prime{\bf Y}^\prime
                  \right]
         \right\}
     \det(x-{\bf X}^\prime)\det(y-{\bf Y}^\prime)
   }
   {\int {\rm d}{\bf X}^\prime{\rm d}{\bf Y}^\prime
     \exp\left\{
                -\frac{N-1}{g^\prime} 
                 {\rm Tr}
                   \left[
                         U({\bf X}^\prime)+\widetilde{U}({\bf Y}^\prime)
                         -{\bf X}^\prime{\bf Y}^\prime
                   \right]
         \right\}
   }\,,
\nonumber\\
\label{effectiveaction}
\end{eqnarray}
where $g^\prime=(1-1/N)g$. 

To obtain $V_{\rm eff}(x,y)$, 
one may calculate the right-hand side of eq.(\ref{effectiveaction}) as
\begin{eqnarray}
 &&\exp
   \Biggl\{
       -\frac{N-1}{g^\prime}
         \left[
               U(x)+\widetilde{U}(y)-x y
         \right]
       +\left\langle{{\rm Tr}\log(x-{\bf X}^\prime)}\right\rangle
       +\left\langle{{\rm Tr}\log(y-{\bf Y}^\prime)}\right\rangle 
       \nonumber\\
 &&\quad\quad\ +\frac{1}{2}
         {\left\langle{\rm Tr}\log(x-{\bf X}^\prime)
                      {\rm Tr}\log(x-{\bf X}^\prime)
         \right\rangle}_{\rm c}
       +\frac{1}{2}
         {\left\langle{\rm Tr}\log(y-{\bf Y}^\prime)
                      {\rm Tr}\log(y-{\bf Y}^\prime)
         \right\rangle}_{\rm c} 
       \nonumber\\
 &&\quad\quad\ +{\left\langle{\rm Tr}\log(x-{\bf X}^\prime)
                             {\rm Tr}\log(y-{\bf Y}^\prime)
        \right\rangle}_{\rm c} 
       +\cdots
   \Biggr\}\,,\label{expansion}
\end{eqnarray}
where
\begin{equation}
 {\left\langle F({\bf X}^\prime,{\bf Y}^\prime) \right\rangle}=
  \frac
   {
    \int {\rm d}{\bf X}^\prime {\rm d}{\bf Y}^\prime 
     \exp
      \left\{
             -\frac{N-1}{g^\prime} 
              {\rm Tr}
                \left[
                      U({\bf X}^\prime)+\widetilde{U}({\bf Y}^\prime)
                      -{\bf X}^\prime{\bf Y}^\prime
                \right]
     \right\}
    F({\bf X}^\prime,{\bf Y}^\prime)
   }
   {
    \int {\rm d}{\bf X}^\prime {\rm d}{\bf Y}^\prime 
     \exp
      \left\{
             -\frac{N-1}{g^\prime} 
               {\rm Tr}
                 \left[
                       U({\bf X}^\prime)+\widetilde{U}({\bf Y}^\prime)
                       -{\bf X}^\prime{\bf Y}^\prime
                 \right]
      \right\}
  }\,,  
\end{equation}
and the subscript ${\rm c}$ denotes the connected part. 
Since insertions of ${\rm Tr} \log(x-{\bf X}^\prime)$ and ${\rm Tr} \log(y-{\bf Y}^\prime)$ correspond to 
boundaries (or loops) on the worldsheet,  eq.(\ref{expansion}) can be
considered as a genus expansion of the free energy in open string theory.

However, this expansion is not always valid. In the one-matrix case, as was discussed in 
\cite{Ishibashi:2005dh}, such an expansion is not valid when the argument of $V_{\rm eff}$ 
is in the region where other eigenvalues are distributed. 
For such an argument, the effect of the repulsive potential 
between two eigenvalues cannot be treated perturbatively in ${1}/{N}$. This argument is 
also valid in the two-matrix case, and we cannot trust the expansion eq.(\ref{expansion}) for 
$x$ or $y$ in the region where other $x$'s or $y$'s are distributed. 
Therefore we should calculate the effect of the eigenvalues in these regions using some 
other techniques. On the other hand, 
as we will see in the next section, 
the nonperturbative effects are due to the isolated saddle points for which $x$ and $y$ are outside 
such regions. Therefore we can use eq.(\ref{expansion}) to study the fluctuations around these saddle 
points. The situation is quite the same as that for the one-matrix model and we can get the nonperturbative 
correction to the free energy due to the saddle point $(x,y)$ as \cite{Hanada:2004im}
\begin{equation}
 \delta F=N\frac{\int_{(x,y)}
{\rm d} x^\prime{\rm d} y^\prime \exp\left[-V_{\rm eff}(x^\prime,y^\prime)
\right]}
{\int
{\rm d} x^\prime{\rm d} y^\prime \exp\left[-V_{\rm eff}(x^\prime,y^\prime)
\right]}\,. \label{correction}
\end{equation}
The numerator corresponds to the configuration
where one set of eigenvalues 
is sitting at the saddle point $(x,y)$. 
If this saddle point survives in the double scaling limit, 
the corresponding configuration 
would be regarded as a D-instanton
configuration.
We can evaluate 
the integral in the numerator of eq.(\ref{correction}) 
around $(x,y)$ by the saddle point method.
In the double scaling limit, $\delta F$ becomes of the form
\begin{equation}
\exp\left[-S_0/g_{\rm s}+\log \mu_{\rm inst.}
+\mathcal{O}(g_{\rm s})\right]\,. \label{action_chemical}
\end{equation}
Here, $S_0$ and $\mu_{\rm inst.}$
can be considered as the classical action and the
chemical potential of the D-instanton, respectively.
The main purpose of this paper is to calculate $\delta F$ in the 
double scaling limit and see if it is a universal quantity.

\section{The two-matrix model}
For the calculation of the numerator of eq.(\ref{correction}), we can use 
eq.(\ref{expansion}).
Using this formula, we obtain the leading order
contribution to $V_{\rm eff}(x,y)$ 
as
\begin{equation}
 V^{(0)}_{\rm eff}(x,y)
 =\frac{N-1}{g^\prime} 
   \left[
         U(x)+\widetilde{U}(y)-x y
   \right]
  -{\left\langle{\rm Tr} \log(x-{\bf X}^\prime)\right\rangle}
  -{\left\langle{\rm Tr} \log(y-{\bf Y}^\prime)\right\rangle}\,. 
\label{vefflargeN}
\end{equation}
We can evaluate this from 
the large-$N$ limit of the resolvents,
\begin{equation}
 W(x)            =\frac{1}{N-1} 
                   {\left\langle{\rm Tr}\frac{1}{x-{\bf X}^\prime}
                    \right\rangle}\,,\
 \widetilde{W}(y)=\frac{1}{N-1} 
                   {\left\langle{\rm Tr}\frac{1}{y-{\bf Y}^\prime}
                    \right\rangle}\,. 
\end{equation} 
For convenience, we rather
consider the following combinations,
\begin{equation}
Y(x)=               U'(x)-{g^\prime}            W(x)\,,\ 
X(y)= {\widetilde{U}}'(y)-{g^\prime}{\widetilde W}(y)\,\label{fct:XY}.
\end{equation}

The functions $Y(x)$ and $X(y)$ have been much studied in the
literature (see \cite{Eynard:2002kg} and references therein).  
Here we summarize some known properties of $Y(x)$ and $X(y)$ 
for later convenience.
Note first that as a function of $x$ ($y$) on the complex plane, 
$Y(x)$ ($X(y)$) has a cut on the real axis, where the eigenvalues of 
${\bf X}^\prime$ (${\bf Y}^\prime$) 
are distributed.
One can define the Riemann surface of
$Y(x)$ ($X(y)$). 
The complex plane we started from is 
called the physical sheet of $x$ ($y$).
In other words, the physical sheet of $x$ ($y$) is 
the sheet that  
contains the infinity $x=\infty_x$ ($y=\infty_y$) around which
$Y(x)$ ($X(y)$) is expanded as
\begin{equation}
       Y(x)\sim 
             U^\prime(x)-\frac{g^\prime}{x}+{\mathcal{O}}(x^{-2})
\quad
\biggl(
       X(y)\sim 
             {\widetilde U}^\prime(y)-\frac{g^\prime}{y}
             +{\mathcal{O}}(y^{-2})
\biggr)\,.
\end{equation}

Actually the functions $Y(x)$ and $X(y)$ satisfy 
\begin{equation}
 y=Y(x)\,,\ x=X(y)\,, \label{Riemann}
\end{equation}
which can be proved by showing that the pairs $(x,Y(x))$ and 
$(X(y),y)$ satisfy the same algebraic equation. 
This relation implies that 
the Riemann surfaces of $Y(x)$ and $X(y)$ are actually the same.
As usual,  
eq.(\ref{Riemann}) is solved for $x, y$  as 
$x={\mathcal X}(s),\ y={\mathcal
Y}(s)$. Here, ${\mathcal X}(s)$ and ${\mathcal Y}(s)$ 
are certain functions of a uniformization parameter $s$,
 which globally parametrizes the Riemann
surface of eq.(\ref{Riemann}).
Since 
we are interested in $(p,q)$ noncritical string theory, 
we can restrict ourselves to 
the Riemann surface with genus zero.
In this case, the uniformization parameter
$s$ takes values in ${C}\cup \infty$, 
and the functions ${\mathcal X}(s)$ and ${\mathcal
Y}(s)$ are known to be expanded as 
\begin{eqnarray}
{\mathcal X}(s)&\!\!\!=\!\!\!&\gamma s
                 +\sum_{k=0}^{\widetilde d-1}\frac{\alpha_k}{s^k}\,,
\nonumber\\
{\mathcal Y}(s)&\!\!\!=\!\!\!&\frac{\gamma}{s}
                 +\sum_{k=0}^{d-1}{\beta_k}{s^k}\,.
\label{expXY}
\end{eqnarray}
Here, the coefficients $\alpha_k$ and $\beta_k$ are determined from 
the potentials $U(x)$ and ${\widetilde{U}}(y)$, and $d$ and $\widetilde
d$ are the degrees of $U(x)$ and $\widetilde{U}(y)$, respectively.
In terms of ${\mathcal X}(s)$ and ${\mathcal Y}(s)$, 
we can express $Y(x)$ and $X(y)$ as
\begin{equation}
 Y(x)={\mathcal Y}({\mathcal X}^{-1}(x))\,,\ 
 X(y)={\mathcal X}({\mathcal Y}^{-1}(y))\,.\label{sol:1}
\end{equation}
Let us note that
\begin{equation}
 \lim_{s\to \infty}{\mathcal X}(s)=\infty_x\,,\ \lim_{s\to 0}{\mathcal Y}(s)=\infty_y\,.
\end{equation}
These equations
 mean that $x=\infty_x$ corresponds to $s=\infty$ and $y=\infty_y$ 
corresponds to $s=0$. 

Once the functions $Y(x)$ and $X(y)$ are given as in eq.(\ref{sol:1}), 
we obtain $V^{(0)}_{\rm eff}(x,y)$ as
\begin{eqnarray}
&&\frac{N-1}{g^\prime}
 \lim_{\Lambda_x \to \infty_x}\lim_{\Lambda_y\to\infty_y}
   \Biggl[
           \int_{\Lambda_x}^x {\rm d}x^\prime Y(x^\prime)
           + \int_{\Lambda_y}^y {\rm d}y^\prime X(y^\prime)
           - x y\nonumber\\
&&\qquad\qquad\qquad +U(\Lambda_x)+U(\Lambda_y)
            -g^\prime \log (\Lambda_x\Lambda_y) 
   \Biggr]\,. \label{veff:0}
\end{eqnarray}
In the following, whenever $\Lambda_x$ or $\Lambda_y$ appears, it is 
understood that we take the limit $\lim_{\Lambda_x\to\infty}$ or 
$\lim_{\Lambda_y\to\infty}$ and we will not write these symbols 
explicitly. 
We can also express $V^{(0)}_{\rm eff}(x,y)$
in terms of the single-valued 
functions ${\mathcal X}(s)$ and ${\mathcal Y}(s)$ 
on the Riemann surface.
Indeed, by changing the variables $x$ and $y$ as
$x={\mathcal X}(s)$ and $y={\mathcal Y}(\tilde{s})$, we obtain
\begin{eqnarray}
V^{(0)}_{\rm eff}(x,y)
&\!\!\!=\!\!\!&
  \frac{N-1}{g^\prime}
   \Biggl[
           \int_{\Lambda_x/\gamma}^s 
             {\rm d}s^\prime 
              {\mathcal Y}(s^\prime)\partial{\mathcal X}(s^\prime)
               +\int_{\gamma/\Lambda_y}^{\tilde{s}} 
             {\rm d}s^\prime 
              {\mathcal X}(s^\prime)\partial{\mathcal Y}(s^\prime)
              -{\mathcal X}(s) {\mathcal Y}(\tilde{s})\nonumber\\
&&\qquad+U(\Lambda_x)+\widetilde{U}(\Lambda_y)
           -{g^\prime}\log(\Lambda_x\Lambda_y)
   \Biggr]\,. \label{veff:1}
\end{eqnarray}

From the expression eq.(\ref{veff:1}), we can see that 
the potential $V^{(0)}_{\rm eff}(x,y)$ 
has local extrema at
$(x,y)$
with $x$ and $y$ 
satisfying the saddle point equations,
\begin{equation}
x= {\mathcal X}(s)={\mathcal X}({\tilde{s}})\,,\ 
y={\mathcal Y}(s)={\mathcal Y}(\tilde{s})\,.\label{dbpt}
\end{equation}
The saddle points solving eq.(\ref{dbpt}) are divided into two classes.
One is of the trivial saddle points $(x,y)$ satisfying $s=\tilde s$.
The other is of the non-trivial saddle points 
$(x,y)$ satisfying $s\neq\tilde s$. 
It is not 
the former class but the latter class of saddle points, 
which can contribute to 
the nonperturbative corrections to the free energy \cite{Kazakov:2004du}.
Thus, we consider only 
the latter class of saddle points below.
The non-trivial saddle points
are nothing but 
the double points, or the singularities, of the Riemann 
surface \cite{Seiberg:2003nm}.

\section{The chemical potential of D-instantons}

In order to get the chemical potential of D-instantons, we should
consider the next to leading order contribution in the large-$N$ limit. 
From eq.(\ref{expansion}) we can see that we need two loop correlators in
the two-matrix model. 
It is known that these correlators
can be expressed in terms of the functions ${\mathcal X}(s),\ {\mathcal
Y}(s)$ as follows \cite{Daul:1993bg}:
\begin{eqnarray}
 {\left\langle
      {\rm Tr}\log(x-{\bf X}^\prime)
      {\rm Tr}\log(x^\prime-{\bf X}^\prime)
  \right\rangle}_c
 &\!\!\!=\!\!\!&
    -\log\frac{{\mathcal X}(s)-{\mathcal X}(s^\prime)}{s-s^\prime}
    +\log\gamma\,,
    {\nonumber}\\
 {\left\langle
      {\rm Tr}\log(y-{\bf Y}^\prime)
      {\rm Tr}\log(y^\prime-{\bf Y}^\prime)
  \right\rangle}_c
 &\!\!\!=\!\!\!&
    -\log\frac
          {
           {\mathcal Y}(\tilde{s})-{\mathcal Y}({\tilde{s}}')
          }
          {1/{\tilde{s}}-1/{\tilde{s}}'
          }
    +\log\gamma\,, 
    {\nonumber}\\
 {\left\langle
      {\rm Tr}\log(x-{\bf X}^\prime)
      {\rm Tr}\log(y-{\bf Y}^\prime)
  \right\rangle}_c
&\!\!\!=\!\!\!&
     -\log\left(1-\frac{\tilde{s}}{{s}}\right)\,.\label{correlators}
\end{eqnarray}
Here, $x, x^\prime, y, y^\prime$ are related to 
$s,s^\prime,{\tilde s},{\tilde
s}^\prime$ as 
$x={\mathcal X}(s),\ x^\prime={\mathcal X}(s^\prime),\ 
y={\mathcal Y}(\tilde{s}),\ y^\prime={\mathcal Y}({\tilde{s}}^\prime)$.

Using these correlators, we can evaluate 
$V_{\rm eff}(x^\prime, y^\prime)$ 
for the point $(x^\prime,y^\prime)
=({\mathcal X}(s^\prime),{\mathcal Y}({\tilde s}^\prime))$ 
near the double point 
$(x,y)
=({\mathcal X}(s),{\mathcal Y}({\tilde s}))$
as
\begin{eqnarray}
V_{\rm eff}(x^\prime,y^\prime)
&\!\!\!=\!\!\!&
     \frac{N-1}{g^\prime} V^{(0)}_{\rm eff}(x^\prime,y^\prime)\nonumber\\ 
& & +\frac{1}{2}
        \log\left[
                  \partial {\mathcal X}(s^\prime)
            \right]
   +\frac{1}{2}
        \log\left[
	          -\tilde{s}^{\prime 2}\partial {\mathcal Y}(\tilde{s}^\prime)
            \right]
   -\log\gamma
   +\log\left(
               1-\frac{\tilde{s}^\prime}{s^\prime}
        \right)\nonumber\\
& & +{\mathcal O}(1/N)\,,\label{veff:2}
\end{eqnarray}
where $V_{\rm eff}^{(0)}(x^\prime,y^\prime)$ 
is the potential calculated from eq.(\ref{veff:1}).
Now, it is straightforward to evaluate the numerator of
eq.(\ref{correction}).  
The result is
\begin{eqnarray}
&&\int_{(x,y)}{\rm d}x^\prime {\rm d}y^\prime
  \exp
   \left[
         -V_{{\rm eff}}(x^\prime,y^\prime)
   \right]\nonumber\\
&&\simeq\gamma
   \left(
         1-\frac{\tilde{s}}
                {s}
   \right)^{-1}
   \tilde{s}^{-1}
   \left[
         -{\partial {\mathcal X}(s)}
          {\partial {\mathcal Y}(\tilde{s})}
   \right]^{-1/2}
   \left(
         \frac{2\pi g^\prime}
              {N-1}
   \right) 
   \left[
         \frac{\partial  {\mathcal Y}(s)}
              {\partial  {\mathcal X}(s)}
         \frac{\partial  {\mathcal{X}}({\tilde{s}})}
              {\partial  {\mathcal Y}(\tilde{s})}
         -1
   \right]^{-1/2}
{\nonumber}\\
&&\qquad\qquad\times
 \exp
  \left\{
         \frac{N-1}
              {g^\prime} 
         \left[
               2{\rm R}
               -\int_{{\tilde{s}}}^{s} {\rm d}s^\prime
                  {\mathcal Y}(s^\prime)\partial  {\mathcal X}(s^\prime)
         \right]
  \right\}\,,
 \label{numerator}
\end{eqnarray}
where
\begin{eqnarray}
2{\rm R}&\!\!\!=\!\!\!&\int_{\gamma/\Lambda_y}^{\Lambda_x/\gamma}
                {\rm d}s^\prime {\mathcal Y}(s^\prime)
                                \partial{\mathcal X}(s^\prime)
                +{\mathcal X}(\gamma/\Lambda_y)
                 {\mathcal Y}(\gamma/\Lambda_y)
                -U(\Lambda_x)-{\widetilde{U}}(\Lambda_y)
                +{g^\prime}\log(\Lambda_x\Lambda_y)
  \nonumber\\
&\!\!\!=\!\!\!&\int_{X(\Lambda_y)}^{\Lambda_x}
                {\rm d}x^\prime Y(x^\prime)
                +X(\Lambda_y)\Lambda_y
                -U(\Lambda_x)-{\widetilde{U}}(\Lambda_y)
                +{g^\prime}\log(\Lambda_x\Lambda_y)\,.\label{R}
\end{eqnarray}

The denominator of eq.(\ref{correction}) can be calculated 
as follows \cite{Ishibashi:2005dh}.
First, let us note that
the denominator of eq.(\ref{correction}) is expressed as
\begin{eqnarray}
&&\hspace{-20pt}
  \int {\rm d}x {\rm d}y 
  \exp
   \left[
         -V_{\rm eff}(x,y)
   \right]
  {\nonumber}\\
&&\hspace{-20pt}
  =\frac
    {
     \int {\rm d}x {\rm d}y 
     \int {\rm d}{\bf X}^\prime {\rm d}{\bf Y}^\prime
      \exp
       \left\{
              -\frac{N-1}{g^\prime} 
               {\rm Tr}
                 \left[
                       U({\bf X}^\prime)+{\widetilde{U}}({\bf Y}^\prime)
                       -{\bf X}^\prime{\bf Y}^\prime
                 \right]
       \right\}
      \det(x-{\bf X}^\prime)\det(y-{\bf Y}^\prime)
    }
    {
     \int {\rm d}{\bf X}^\prime{\rm d}{\bf Y}^\prime
      \exp
       \left\{
              -\frac{N-1}{g^\prime} 
               {\rm Tr}
                 \left[
                       U({\bf X}^\prime)+{\widetilde{U}}({\bf Y}^\prime)
                       -{\bf X}^\prime{\bf Y}^\prime
                 \right]
      \right\}
     }\,.
{\nonumber}\\
&& \label{denominator:0}
\end{eqnarray}
Since the numerator of eq.(\ref{denominator:0})
is proportional to the partition function
\begin{equation}
 \int {\rm d}{\bf X} {\rm d}{\bf Y} 
  \exp
   \left
    \{
      -\frac{N}{g}{\rm Tr} 
        \left[
              U({\bf X})+{\widetilde{U}}({\bf Y})-{\bf X}{\bf Y}
        \right]
    \right\}\,,
\end{equation}
what we should calculate is essentially the ratio between the matrix
integrals over $({\bf X},{\bf Y})$ and over $({\bf X}^\prime,{\bf Y}^\prime)$. After some
calculations, which are presented in the appendix B, 
we obtain
\begin{equation}
\int {\rm d}x {\rm d}y
 \exp
  \left[
        -V_{{\rm eff}}(x,y)
  \right]
\simeq
 (2\pi)^{\frac{3}{2}}
 \gamma 
 \sqrt{(N-1)g^\prime}
 \exp
  \left[
        \frac{2(N-1){\rm R}}{g^\prime}
  \right]
\,. \label{denominator}
\end{equation}

From eq.(\ref{numerator}) and eq.(\ref{denominator}), 
we get
\begin{eqnarray}
\delta F&\!\!\!\simeq\!\!\!&
          \sqrt{
                \frac{g^\prime}{2\pi(N-1)}
               }
          \frac{
                \left(
                      1-{\tilde{s}}/{s}
                \right)^{-1}
                \tilde{s}^{-1}
               }
               {
                \left[
                      \partial  {\mathcal X}(s)
                      \partial  {\mathcal Y}(\tilde{s})
                     -\partial  {\mathcal X}(\tilde{s})
                      \partial  {\mathcal Y}(s)
                \right]^{1/2}
               }
          \exp
           \left[
                 -\frac{N-1}{g^\prime}
                  \int_{\tilde{s}}^s {\rm d}s^\prime
                   {\mathcal Y}(s^\prime)\partial{\mathcal X}(s^\prime)
           \right]\,.
\nonumber\\
&&\label{correction:2}
\end{eqnarray}
It is possible to express eq.(\ref{correction:2}) in terms of 
${\mathcal X}(s)$ and $Y(x)={\mathcal Y}({\mathcal X}^{-1}(x))$.
By doing so, one
finds that $\delta F$ depends not on $Y(x)$ itself
but on the non-polynomial part, or the singular part,
of $Y(x)$. Next, we will use this fact 
to evaluate $\delta F$ in the double scaling
limit. 

We need to know how ${\mathcal X}(s)$ and the singular part of $Y(x)$
scale as we take the double scaling limit which realizes $(p,q)$
noncritical string theory.
Such a scaling limit is described in the following 
way \cite{Daul:1993bg}.
Let us 
introduce a positive cut-off parameter $a\propto
N^{-1/(p+q)}$ and consider the limit $a\to 0\ (N\to \infty)$.
In order to describe the limit, 
let us parametrize $s$ by a new uniformization
parameter $\omega$ as $s=s_* \exp(a \xi \omega)$, 
where $s_*$ is the would-be critical point and $\xi$ is a positive
parameter, whose power $\xi^{2p}$ should be identified with the cosmological constant. 
Now, we take the limit $a\to 0$ while we tune 
the potentials $U(x)$ and ${\widetilde
U}(y)$ to realize $(p,q)$ noncritical string theory. 
Such a fine tuning can be carried out 
so that the leading order contribution to 
$x={\mathcal X}(s)$ and the singular part of $Y(x)$ are given as
\begin{eqnarray}
x-x_*&\!\!\!\sim\!\!\!&
2 {C} a^p \xi^p {\rm T}_p(\omega)\,,\nonumber\\
Y_{\rm sing}(x)-Y_{\rm sing}(x_*)
&\!\!\!\sim\!\!\!& 
2 \widetilde{C} a^q \xi^q {\rm T}_q(\omega)\,,\label{cheb:1}
\end{eqnarray}
where $x_*={\mathcal X}(s_*)$ and 
${\rm T}_n(\cosh\theta)=\cosh(n\theta)$ is the Chebyshev
polynomial of the first kind.
Here, $C$ and $\widetilde C$ are certain constants.

Let us introduce new scaling variables $\hat x$ and $\hat y$ as
\begin{equation}
 x-x_*=2C a^p \xi^p \hat x\,,\  Y_{\rm sing}(x)-Y_{\rm sing}(x_*)=2{\widetilde C} a^q \xi^q \hat y\,.
\end{equation}
Then, 
we can see that in the double scaling limit, the Riemann surface of
eq.(\ref{Riemann}) is reduced to that of
\begin{equation}
 {\rm T}_q(\hat x)-{\rm T}_p(\hat y)=0\ .\label{Riemann:2}
\end{equation}
In this way, we reach the Riemann surface associated with $(p,q)$
noncritical string \cite{Seiberg:2003nm}.
As was shown in \cite{Seiberg:2003nm}, 
the Riemann surface of (\ref{Riemann:2}) is again with genus zero and has
$(p-1)(q-1)/2$ inequivalent double points.
To describe the Riemann surface of eq.(\ref{Riemann:2}) and its double points, 
let us parametrize $\omega$ as $\omega=\cosh\theta$.
In this parametrization, we have
\begin{equation}
\hat x=\cosh(p\theta)\,,\ \hat y=\cosh(q\theta)\,. 
\end{equation} 
Thus, we can conclude
that in the double scaling limit, the double points are given by the pairs,  
\begin{equation}
(\theta_{m,-n},\theta_{m,n})\,, 
\end{equation}
where
\begin{equation}
 \theta_{m,n}=i\pi\left(\frac{m}{p}+\frac{n}{q}\right)\,,
\end{equation}
with $1\leq m\leq p-1,\ 1\leq n\leq q-1$ and $m q-n p>0$ \cite{Seiberg:2003nm}.

Now 
it is not difficult to evaluate $\delta F$ in the double scaling limit.
By putting all ingredients together
and taking $N\to \infty, a\to 0$ with 
\begin{equation}
N |C \widetilde{C}| g^{-1} a^{p+q}=g_{\rm s}^{-1}\ , 
\end{equation}
fixed, we obtain
\begin{eqnarray}
 \delta F&\!\!\!=\!\!\!&\frac{1}{8}
            \sqrt{
                  \frac{g_{\rm s}}{2\pi p q \xi^{p+q}}
                 }
            \left(
                  \sin\frac{\pi m}{p} \sin\frac{\pi n}{q}
            \right)^{-1}
            \nonumber\\
          &&\qquad\times 
           \left(
                 \cos\frac{2\pi m}{p}-\cos\frac{2\pi n}{q}
           \right)^{1/2}
           \left[
                 \eta (-1)^{m+n}
                 \sin\frac{\pi m q}{p} \sin\frac{\pi n p}{q}
           \right]^{-1/2}
\nonumber\\
          &&\qquad\qquad\times
           \exp
            \left[
                  -\eta (-1)^{m+n}\frac{8 p q }{g_{\rm s}(q^2-p^2)} 
                   \xi^{p+q}
                   \sin\frac{\pi m q}{p} \sin\frac{\pi n p}{q}
            \right]\,. \label{correction:3}
\end{eqnarray}
Here $\eta$ is 
the sign of $C\widetilde C$, 
which can be
 determined as $\eta={\rm sign}(\sin({\pi q}/{p}))$.
This follows from the condition that the distribution 
function of eigenvalues of 
${\bf X}^\prime$, defined by
\begin{equation}
\rho_x(x)=-\frac{1}{\pi}{\rm Im}\,W(x+{\rm i 0})
=\frac{1}{\pi g^\prime}{\rm Im}\,Y(x+{\rm i} 0)\,, 
\end{equation}
satisfies $\rho_x(x)\geq 0$. Here, $x$ was understood to be 
on the physical sheet\footnote{In the parametrization $x\propto \cosh p\theta$,
the physical sheet of $x$ is the domain 
of $0<|{\rm Im} \theta|<\pi/p$ \cite{Kazakov:2004du}.}.  

Thus we have obtained $\delta F$ in the double scaling limit. Essentially it depends 
only on the combination ${\xi^{p+q}}/{g_{\rm s}}$ which is the scaling variable in the 
continuum theory. Therefore we have shown that $\delta F$ to this order is a universal 
quantity. As a check of the universality, we can compare the result with that of the one 
matrix model. One can realize the $(2,2k+1)$ noncritical string theory as the higher critical 
point of the one-matrix model. It is possible to calculate $\delta F$ for such string theory 
using the results in \cite{Ishibashi:2005dh}. As we will show in appendix A, the result perfectly 
agrees with eq.(\ref{correction:3}).

Here we have identified the double points $(s,\tilde s)$ in eq.(\ref{correction:2}) 
with $(\theta_{m,-n},\theta_{m,n})$ via $s=s_* \exp(a\xi\cosh\theta)$.
In principle, one can also consider the double points $(\theta_{m,n},\theta_{m,-n})$, 
whose classical action $S_0$ has the sign opposite to the one in eq.(\ref{correction:3}). 
Our choice is a natural one such that the action part $S_0$  
agrees with the one proposed in the continuum theory \cite{Kutasov:2004fg}, 
with $\eta$ given above.  

\section{Conclusions and discussions}
In this paper we generalize the result in \cite{Hanada:2004im} 
to the two-matrix model case and study the nonperturbative effects 
for $(p,q)$ noncritical string theory. 
Utilizing the method 
given in \cite{Ishibashi:2005dh}, 
we can obtain the nonperturbative effects in the form of 
$\exp [-S_0/g_{\rm s}+\log \mu_{\rm inst}]$. 
We find that $S_0$ reproduces the known results and that 
$\mu_{\rm inst}$ is finite and universal as in the case of the $c=0$ noncritical 
string theory. Although there are many non-universal parameters involved, 
the final result depends only on the scaling variable
${\xi^{p+q}}/{g_{\rm s}}$. 

Although we get the number, the physical meaning of $\delta F$ is not obvious for 
most of $(p,q)$ and $(m,n)$. For some combination $\delta F$ becomes imaginary and 
we can see that it is related to some instability. However for the two-matrix model, 
we know the behavior of the effective action $V_{\rm eff}(x,y)$ only around the saddle points, 
and we cannot get the clear picture of such instabilities.

It will be intriguing to study what we have obtained from the point of view 
of the loop equation or the string field theory. 
In \cite{Hanada:2004im}, the nonperturbative effects of $c=0$ string theory 
were studied from this point of view, although the calculation of the chemical 
potential itself seems difficult in such an approach. 
In a recent paper \cite{Fukuma:2005nm}, the chemical potential for $(p,p+1)$ noncritical 
string theory was calculated by making some assumptions in the SFT approach 
proposed in \cite{Fukuma:1996hj}.
Their results agree with ours eq.(\ref{correction:3}) up to a factor of $i$. 
It would be interesting to check if we can perform similar calculations using 
other string field theories \cite{Ishibashi:1993sv,Ikehara:1994vx}. 

It would also be interesting to compute loop amplitudes 
in a fixed D-instanton background as was done in \cite{Hanada:2004im} 
for the $c=0$ case. 
Comparing it with the Liouville results, 
we will be able to get a strong evidence 
for identifying the isolated eigenvalue  
as the D-instanton in $c<1$ noncritical string theory. 
Such a study will be useful to understand what D-brane is in string theory.   

We believe our work would be 
instructive to the study of nonperturbative effects of critical string theory. 
The proposed candidates of the nonperturbative formulation of critical string theory 
are, in sense, multi-matrix models 
\cite{Banks:1996vh,Ishibashi:1996xs,Maldacena:1997re}. 
Thus in order to 
go from $c=0$ to critical string theory, it is indispensable to know 
how to extend the analysis in the case of the one-matrix model 
to a multi-matrix model.   

\vspace{0.5cm}
\noindent
{\Large \bf Acknowledgments} \\
The authors would like to thank H. Kawai, I. Kostov, T. Matsuo, and 
S. Yahikozawa for valuable discussions.
This work is supported in part by the Grants-in-Aid for Scientific Research 
13135224 and 13640308. The work of T.K. is supported in part by the Special 
Postdoctoral Researchers Program.

\appendix
\section{The $(2,2k+1)$ models}
The nonperturbative effects for the higher critical points of the 
one-matrix model can be studied in the same way as in the $c=0$ 
case\footnote{See \cite{Sato:2004tz}, 
for the analysis generalizing 
\cite{Hanada:2004im}.}.

In the one-matrix model, the resolvent can be obtained as
\begin{equation}
{\left\langle \frac{1}{N}{\rm Tr}\frac{1}{z-M}\right\rangle}
=\frac{1}{2 g^2}\left[
V'(z)-M(z)\sqrt{(z-\alpha)(z-\beta)}
\right]\ ,
\end{equation}
where $M(z)$ is a polynomial.
In the continuum limit corresponding to the $(2,2k+1)$ model, we take
$a\to 0$ with
\begin{eqnarray}
 z &\!\!\!\sim\!\!\!& \beta+a[\cosh(2\theta)+1]\ ,\nonumber\\
 \frac{1}{g^2}M(z)\sqrt{(z-\alpha)(z-\beta)}
&\!\!\!\sim\!\!\!& C(-1)^k a^{k+1/2}\cosh[(2k+1)\theta]\sqrt{\beta-\alpha}\ ,
\label{higher}
\end{eqnarray}
where $C$ is a positive constant.
Using eq.(\ref{higher}),
we can evaluate $V^{(0)}_{\rm eff}(z)$, the large-$N$ limit of 
the effective potential of the matrix eigenvalues 
\cite{Hanada:2004im} and
see that the isolated saddle points of $V^{(0)}_{\rm eff}(z)$ 
are located at  
the positions corresponding to
\begin{equation}
\theta=\frac{\pi{\rm i}(2 l+1)}{2(2k+1)}\,, 
\end{equation}
with $l=0,1,\cdots,k-1$.
Then, using the results in \cite{Ishibashi:2005dh}, 
we can show that these saddle points 
contribute to the nonperturbative corrections to the free energy as
\begin{eqnarray}
\delta F&\!\!\!=\!\!\!& \frac{t^{-(2k+3)/8}}{4\sqrt{\pi}\left[1+\cos(\frac{2l+1}{2k+1}\pi)\right]}
\left[(-1)^{l+k}\frac{\sin\left(\frac{2l+1}{2k+1}\pi\right)}{2k+1}\right]^{1/2}
\nonumber\\
&&\qquad\times \exp\left[(-1)^{l+k}t^{(2k+3)/4}\frac{2(2k+1)}{(2k+1)^2-4}\sin\left(
\frac{2l+1}{2k+1}\pi\right)\right]\ , \label{chem:higher}
\end{eqnarray}
with $t$ the cosmological constant.
We can see
that eq.(\ref{chem:higher}) indeed agrees with 
eq.(\ref{correction:3}) by 
choosing $(m,n)=(1,k-l),\ g_{\rm s}=8$ and 
$\xi=t^{1/4}$ in eq.(\ref{correction:3}). 

\section{The denominator}
In this appendix, we will explain how to calculate $\int {\rm d} x{\rm d}
y\exp[-V_{\rm eff}(x,y)]$ and get the result eq.(\ref{denominator}) in the large-$N$
limit.

For this purpose, it is convenient to define $Z_N$ and $Z^\prime_{N-1}$ as 
\begin{eqnarray}
Z_N
&\!\!\!=\!\!\!&
\int \prod_{i=1}^N {\rm d} x_i{\rm d} y_i
\triangle (x) \triangle (y)
\exp \left\{-\frac{N}{g}\sum_i \left[U(x_i)+{\widetilde{U}}(y_i)-x_i y_i\right])
\right\}\,,
{\nonumber}
\\
Z^\prime_{N-1}
&\!\!\!=\!\!\!&
\int \prod_{i=1}^{N-1} {\rm d} x^\prime_i{\rm d} y^\prime_i
\triangle (x^\prime ) \triangle (y^\prime )
\exp \left\{
-\frac{N-1}{g^\prime}\sum_i
\left[
U(x^\prime_i)+{\widetilde{U}}(y^\prime_i)-x^\prime_i y^\prime_i)
\right]
\right\}\,,
\end{eqnarray}
where $g=g^\prime (1+\frac{1}{N-1})$. Then, 
we can express $\int {\rm d} x{\rm d} y \exp[-V_{eff}(x,y)]$ as $Z_N/Z^\prime_{N-1}$.

$Z_N$ and $Z^\prime_{N-1}$ can be expressed as matrix integrals as 
\begin{eqnarray}
Z_N&\!\!\!=\!\!\!&
 C_{N}
\int {\rm d}{\bf X} {\rm d}{\bf Y}
\exp 
\left\{
-\frac{N}{g}
{\rm Tr}\left[
U({\bf X})+{\widetilde{U}}({\bf Y})-{\bf X} {\bf Y} 
\right]
\right\}\ ,
\\
Z_{N-1}'&\!\!\!=\!\!\!&
 C_{N-1}'
\int {\rm d}{\bf X}^\prime {\rm d}{\bf Y}^\prime
\exp 
\left\{
-\frac{N-1}{g^\prime }
{\rm Tr}\left[
U({\bf X}^\prime)+{\widetilde{U}}({\bf Y}^\prime)-{\bf X}^\prime {\bf Y}^\prime 
\right]
\right\}\ ,
\end{eqnarray}
where $X, Y$ are $N\times N$ matrices and $X^\prime ,Y^\prime$ are $(N-1)\times (N-1)$ matrices 
and $C_N$ and $C^\prime_{N-1}$ are constants. 
Let us assume that the quadratic part of $U(x)$ and ${\widetilde{U}}(y)$ are 
$\frac{c_1}{2}x^2$ and $\frac{c_2}{2}y^2$ respectively and take the
integration measure ${\rm d}{\bf X}{\rm d}{\bf Y}$ and ${\rm d}{\bf X}^\prime{\rm d}{\bf Y}^\prime $ so that 
\begin{eqnarray}
1
&\!\!\!=\!\!\!&
\int {\rm d}{\bf X}{\rm d}{\bf Y}
\exp \left[
-\frac{N}{g}
{\rm Tr}\left(
\frac{c_1}{2}{\bf X}^2+\frac{c_2}{2}{\bf Y}^2-{\bf X}{\bf Y}
\right)
\right]\ ,
{\nonumber}
\\
1
&\!\!\!=\!\!\!&
\int {\rm d}{\bf X}^\prime{\rm d}{\bf Y}^\prime
\exp \left[
-\frac{N-1}{g^\prime }
{\rm Tr}
\left(
\frac{c_1}{2}{{\bf X}^\prime}^{2}+\frac{c_2}{2}{{\bf Y}^\prime}^{2}
-{\bf X}^\prime {\bf Y}^\prime 
\right)
\right]\ .
\end{eqnarray}
Here we assume $c_1c_2-1\neq 0$. 
The constants $C_N$ and $C^\prime_{N-1}$ are determined
by choosing the 
Gaussian potentials
$U(x)=\frac{c_1}{2}x^2, {\widetilde{U}}(y)=\frac{c_2}{2}y^2$.
By using the method of orthogonal polynomials, one can show 
that these constants satisfy
\begin{equation}
\frac{C_N}{C^\prime_{N-1}}\sim
(2\pi )^{\frac{3}{2}}\sqrt{N-1}e^{-(N-1)}
\frac{(g^\prime )^N}{(c_1c_2-1)^{N-\frac{1}{2}}},
\end{equation}
for large $N$.

Now we would like to evaluate $Z_N/Z^\prime_{N-1}$. For large $N$, 
the matrix integral in eq.(\ref{matrixintegral}) can be written as 
\begin{eqnarray}
&&\hspace{-20pt}\int {\rm d}{\bf X}{\rm d}{\bf Y}
\exp \left\{
-\frac{N}{g}{\rm Tr}
\left[
U({\bf X})+{\widetilde{U}}({\bf Y})-{\bf X}{\bf Y}
\right]
\right\}\nonumber\\
&\!\!\!=\!\!\!&
\exp \left[
\frac{N^2}{g^2}F_0(g)+F_1(g)+\frac{g^2}{N^2}F_2(g)+\cdots 
\right]\ .
\label{matrixintegral2}
\end{eqnarray}
Thus, we obtain
\begin{eqnarray}
\frac{Z_N}{Z^\prime_{N-1}}
&\!\!\!=\!\!\!&
\frac{
C_{N}
\int {\rm d}{\bf X} {\rm d}{\bf Y}
\exp 
\left\{
-\frac{N}{g}
{\rm Tr}\left[
U({\bf X})+{\widetilde{U}}({\bf Y})-{\bf X} {\bf Y} 
\right]
\right\}\ 
}
{
C_{N-1}'
\int {\rm d}{\bf X}^\prime {\rm d}{\bf Y}^\prime
\exp 
\left\{
-\frac{N-1}{g^\prime}
{\rm Tr}\left[
U({\bf X}^\prime)+{\widetilde{U}}({\bf Y}^\prime)-{\bf X}^\prime {\bf Y}^\prime 
\right]
\right\}\ 
}
{\nonumber}
\\
&\!\!\!=\!\!\!&
\frac{C_N}{C^\prime_{N-1}}
\exp \left[
\frac{N-1}{g^\prime}\partial
 _{g^\prime}F_0(g^\prime)+\frac{1}{2}\partial _{g^\prime}^2
 F_0(g^\prime)+{\mathcal O}(1/N)
\right]\ .
\label{derivatives}
\end{eqnarray}
Therefore, to determine the denominator of eq.(\ref{correction}), we
need the derivatives of $F_0(g^\prime)$.
$F_0(g^\prime)$ was calculated in \cite{Bertola:2003rp,Eynard:1997jf}. 
Here we will follow their method to get the derivatives of $F_0(g^\prime)$. 

Let us consider 
what happens to $\partial_{g^\prime} F_0(g^\prime)$ if we 
change the potential $U$ 
as $U\to U+\delta U$ where $\delta U$ does
 not involve the quadratic part.
From eq.(\ref{matrixintegral2}), we can show that 
the variation of $F_0(g^\prime)$ itself is calculated as 
\begin{eqnarray}
 \delta F_0(g^\prime)
 &\!\!\!=\!\!\!&\frac{g^\prime}{N-1} 
    \oint_{\infty_x} \frac{{\rm d} x}{2\pi {\rm i}}
     \delta U(x)
     {\left\langle{\rm Tr}\frac{1}{x-{\bf X}^\prime}\right\rangle}
\nonumber \\
&\!\!\!=\!\!\!&-\oint_{\infty_x} 
    \frac{{\rm d} x}{2\pi {\rm i}} \delta U(x)Y(x)\,.
\label{variation}
\end{eqnarray}
Thus, we obtain 
\begin{equation}
\delta \left[\partial_{g^\prime}F_0(g^\prime )\right]
=
-\oint_{\infty_x} \frac{{\rm d} x}{2\pi i}
 \delta U(x)\partial_{g^\prime}Y(x)\,.
\end{equation}
Let us notice that under the integral sign over $x$, $\partial_{g^\prime} Y(x)$
means the derivative of $Y$ with $x$ fixed. We can indicate this by
putting the subscript as $(\partial_{g^\prime} Y)_x$. 
$(\partial_{g^\prime} Y)_x$ is related to 
the derivative $(\partial_{g^\prime} {\mathcal Y})_s$ with the uniformization
parameter $s$ fixed.
Indeed, by setting $x={\mathcal X}(s)$, 
we obtain
\begin{equation}
(\partial_{g^\prime} Y)_x
=
(\partial_{g^\prime} {\mathcal Y})_s-\frac{\partial Y}{\partial x}(\partial_{g^\prime} {\mathcal X})_s\,.
\end{equation}
Similarly, by setting $y={\mathcal Y}(s)$, we obtain 
\begin{equation}
(\partial_{g^\prime} X)_y
=
(\partial_{g^\prime} {\mathcal X})_s-\frac{\partial X}{\partial y}
(\partial_{g^\prime} {\mathcal Y})_s\,. 
\end{equation}
Using these identities, we obtain
\begin{equation}
{\rm d} x(\partial_{g^\prime} Y)_x=-{\rm d}y(\partial_{g^\prime} X)_y\,,
\end{equation}
for $(x,y)=({\mathcal X}(s),{\mathcal Y}(s))$.
Since the singularities of the one form 
${\rm d} x\,\partial_{g^\prime}Y(x)=-{\rm d} y\,\partial_{g^\prime}X(y)$ 
are only the poles at $x=\infty_x $ and $y=\infty_y$,  
it can be identified with ${{\rm d} s}/{s}$. 
Therefore, we get 
\begin{equation}
\delta \left[\partial_{g^\prime}F_0(g^\prime )\right]
=
-[\delta U(x)]_0\,,
\end{equation}
where $[\delta U(x)]_0$ is the coefficient of $s^0$ when we expand 
$\delta U(x)|_{x={\mathcal X}(s)}$ near $s=\infty$, or $s=0$. 
Now from the expression of $Y(x)$, $[\delta U(x)]_0$ can be rewritten as
\begin{equation}
[\delta U(x)]_0
=
\left[
\int^x_{\Lambda_x}{\rm d} x^\prime \delta Y(x^\prime )\right]_0
+\delta U(\Lambda_x)\,,
\end{equation}
or since $\delta X(y)\sim y^{-2}$ for $y\sim \infty_y$, 
\begin{equation}
[\delta U(x)]_0
=\delta
\left[
\int^{X(\Lambda_y)}_{\Lambda_x}{\rm d} x^\prime Y(x^\prime )
\right]
+\delta U(\Lambda_x)\,.
\end{equation}
On the other hand, if we consider the variation of 
the quantity 2{\rm R} (see eq.(\ref{R})) 
under 
$U\rightarrow U+\delta U$,
we obtain
\begin{equation}
\delta
(2{\rm R})
=
\delta\left[
\int^{\Lambda_x}_{X(\Lambda_y)}{\rm d} x^\prime Y(x^\prime )
\right]
-\delta U(\Lambda_x)\,.
\end{equation}
Therefore it coincides with 
$\delta [\partial_{g^\prime}F_0(g^\prime )]$. 
Since we can treat the variation of ${\widetilde{U}}$ in the same way, we can conclude 
that $2{\rm R}$ coincides with 
$\partial_{g^\prime}F_0(g^\prime )$ 
up to some function $f^\prime (g^\prime ,c_1,c_2)$.

We can determine $f^\prime$ by considering the Gaussian case. 
For the Gaussian case, using 
\begin{eqnarray}
{\mathcal X}(s)
&\!\!\!=\!\!\!&
\sqrt{\frac{g^\prime }{c_1c_2-1}}\left(s+\frac{c_2}{s}\right),
\nonumber
\\
 {\mathcal Y}(s)
&\!\!\!=\!\!\!&
\sqrt{\frac{g^\prime}{c_1c_2-1}}\left(\frac{1}{s}+c_1s\right),
\end{eqnarray}
we get 
\begin{equation}
\int^{\Lambda_x}_{X(\Lambda_y)}{\rm d} x^\prime Y(x^\prime )
-U(\Lambda_x)-{\widetilde{U}}(\Lambda_y)+X(\Lambda_y)\Lambda_y
+g^\prime \ln \Lambda_x\Lambda_y
=
-g^\prime\left(\ln\frac{c_1c_2-1}{g^\prime}+1\right).\nonumber\\ 
\end{equation}
Since $\partial_{g^\prime}F_0(g^\prime)$ should vanish 
for the Gaussian case, we obtain
\begin{equation}
\partial_{g^\prime}F_0(g^\prime)
=
2{\rm R}+g^\prime\left(\ln\frac{c_1c_2-1}{g^\prime }+1\right).
\end{equation}

We can express $\partial_{g^\prime}^2F_0(g^\prime  )$ by using
$\gamma$ in eq.(\ref{expXY}). 
To show this, 
we further reduce the size of the
matrices and define
\begin{equation}
Z^{\prime\prime}_{N-2}
=
C^{\prime\prime}_{N-2}
\int {\rm d} {\bf X}^{\prime\prime}{\rm d} {\bf Y}^{\prime\prime}
\exp \left\{
-\frac{N}{g}{\rm Tr}\left[
U({\bf X}^{\prime\prime})+{\widetilde{U}}({\bf Y}^{\prime\prime})
       -{\bf X}^{\prime\prime}{\bf Y}^{\prime\prime}\right]
\right\},
\end{equation}
for $(N-2)\times (N-2)$ matrices
${\bf X}^{\prime\prime},{\bf Y}^{\prime\prime}$. 
Here $C^{\prime\prime}_{N-2}$ is
defined by 
\begin{eqnarray}
&&\hspace{-20pt}
 C^{\prime\prime}_{N-2}
\int {\rm d}{\bf X}^{\prime\prime} {\rm d}{\bf Y}^{\prime\prime} 
\exp \left\{
-\frac{N}{g}
{\rm Tr}\left[
U({\bf X}^{\prime\prime} )+{\widetilde{U}}({\bf Y}^{\prime\prime} )
   -{\bf X}^{\prime\prime} {\bf Y}^{\prime\prime} 
\right]
\right\}
{\nonumber}
\\
&\!\!\!\!=\!\!\!\!&
\int \prod_{i=1}^{N-2} {\rm d} x^{\prime\prime}_i{\rm d} y^{\prime\prime}_i
\triangle (x^{\prime\prime} ) \triangle (y^{\prime\prime} )
\exp \left\{
-\frac{N}{g}\sum_i 
\left[
U(x^{\prime\prime}_i)+{\widetilde{U}}(y^{\prime\prime}_i)
-x^{\prime\prime}_i y^{\prime\prime}_i
\right]
\right\}\ .
\end{eqnarray}
Then we can express ${Z_N Z_{N-2}''}/{(Z_{N-1}')^2}$ in the large-$N$
limit as 
\begin{equation}
\frac{g^\prime}{c_1c_2-1}
\exp \left[
\partial _{g^\prime}^2 F_0(g^\prime)\right]\ .
\end{equation}
On the other hand, ${Z_N Z_{N-2}''}/({Z_{N-1}')^2}$ can be evaluated by
the orthogonal polynomial technique and can be expressed in 
the large-$N$ limit as $\gamma^2$. Therefore we get 
\begin{equation}
\exp \left[
\frac{1}{2}\partial _{g^\prime}^2 F_0(g^\prime)
\right]
=
\gamma\sqrt{\frac{c_1c_2-1}{g^\prime}}\ .
\end{equation}

Putting all these together we obtain
\begin{equation}
\frac{Z_N}{Z^\prime_{N-1}}
=
(2\pi )^{\frac{3}{2}}\gamma
\sqrt{(N-1)g^\prime}
\exp\left[\frac{2(N-1){\rm R}}{g^\prime}\right]\ .
\end{equation}

\end{document}